\documentstyle[11pt]{article}
\begin{document}
\title{\bf Quantisation of  2D-gravity with Weyl and area-preserving
diffeomorphism invariances}
\author{{J.-G. Zhou${}^{\rm a}$,Y.-G. Miao${}^{\rm b}$,J.-Q.Liang${}^{\rm a}$,H.
J.W.M$\ddot{\rm u}$ller-Kirsten${}^{\rm a}$ and Zhenjiu
Zhang${}^{\rm c}$}\\
{\small ${}^{\rm a}$ Department of Physics, University of Kaiserslautern,
P.O.Box 3049,}\\
{\small D-67653 Kaiserslautern, Germany}\\
{\small ${}^{\rm b}$ Institute of Theoretical Physics, University of Amsterdam,}\\
{\small NL-1018 XE Amsterdam, The Netherlands}\\
{\small ${}^{\rm c}$ Center for Relativity Studies, Department of Physics},\\
{\small Huazhong Normal University, Wuhan 430070,}\\
{\small People's Republic of China}}
\date{}
\maketitle
\vskip 48pt
\begin{center}{\bf Abstract}\end{center}
\baselineskip 22pt
The constraint structure of the induced 2D-gravity with the Weyl and
area-preserving diffeomorphism invariances is analysed in the ADM
formulation. It is found that when the area-preserving diffeomorphism
constraints are kept, the usual conformal gauge does not exist, whereas there
is the possibility to choose the so-called ``quasi-light-cone'' gauge, in which
besides the area-preserving diffeomorphism invariance, the reduced Lagrangian
also possesses the SL(2,R) residual symmetry. This observation indicates that
the claimed
correspondence between the SL(2,R) residual symmetry and the area-preserving
diffeomorphism invariance in both regularisation approaches does not hold. The
string-like approach is then applied to quantise this model, but a fictitious non-zero
central charge in the Virasoro algebra appears. When a set of
gauge-independent SL(2,R) current-like fields is introduced instead of the
string-like variables, a consistent quantum theory is obtained, which
means that the area-preserving diffeomorphism invariance can be maintained at
the quantum level.
\newpage
\baselineskip 22pt
The induced 2D-gravity with Weyl and area-preserving diffeomorphism
invariances has attracted much attention recently [1-7]. As is well known, in
the conventional regularisation approach to 2D gravity [8-11], the
diffeomorphism invariance is preserved, while the Weyl invariance is lost. In
the path integral formulation, this can be accomplished by choosing the
diffeomorphism-invariant, but not Weyl-invariant measures for the functional
integrations [10,11]. Nevertheless, one can adopt an alternative regularization
approach in which part of the diffeomorphism invariance is sacrificed so as to
obtain a Weyl-invariant theory. This alternative approach is motivated by the
observation that the Lagrangian, invariant classically with respect to
reparametrization and Weyl transformation, depends on the metric only through
the Weyl-invariant combination [1-3]. From the conformal geometry's point of
view, this idea is based on the fact that the amount of gauge degrees of
freedom provided by the diffeomorphism group $Diff{\cal M}$
of manifold ${\cal M}$ is equivalent
to the amount provided by the combination of the Weyl rescaling and the
area-preserving diffeomorphism subgroup $SDiff{\cal M}$ of $Diff{\cal M}$ [7].
Recently, the
anomalous Ward identities and part of the constraint structure for the induced
2D-gravity with Weyl and area-preserving diffeomorphism invariances have been
discussed in Ref.[3]. Especially, some physical results associated with
area-preserving diffeomorphism invariance, like 2D Hawking radiation [12], have
been studied in Refs.[4,5]. However, there are still some questions that need
to be answered. For example, in the induced 2D-gravity with reparametrization
invariance, when the diffeomorphism constraints are kept, i.e., no gauge
fixings are chosen for them, one has the freedom to choose the conformal
or light-cone gauge. A natural question arises, in the induced 2D-gravity
with area-preserving diffeomorphism invariance, when the Virasoro generators
are required to annihilate the physical states, whether there is the
possibility to pick up the above two gauges. On the other hand, the effective
Lagrangian in the present case has Weyl and area-preserving diffeomorphism
invariance, so that one may ask whether the latter invariance is broken or not at
the quantum level, and supposing it is lost, whether there exists a consistent
approach to quantise this theory, that is, whether the area-preserving
diffeomorphism invariance can be maintained at the quantum level.

In the present paper, the ADM formulation [13-16] is applied to analyse the
constraint structure for the induced 2D-gravity with the Weyl and
area-preserving diffeomorphism invariances. It is shown that when the
area-preserving diffeomorphism constraints are kept, the usual conformal gauge
does not exist, whereas there is the possibility to choose the so-called
``quasi-light-cone'' gauge. It is first found that in the ``quasi-light-cone''
gauge, there is an SL(2,R) residual symmetry in the reduced Lagrangian. Even
though the SL(2,R) currents manifest themselves as generators of the residual
symmetry in the ``quasi-light-cone'' gauge, we find that these currents can be
defined in a gauge-independent way. As indicated in Ref.[1], the
Weyl-invariant approach has some resemblance with Polyakov's light-cone
approach [9] where there is an SL(2,R) residual symmetry which was thought to
be analogous to the area-preserving group in the Weyl-invariant approach.
However, it is found that in the Weyl-invariant regularisation approach,
besides the area-preserving diffeomorphism invariance, the reduced Lagrangian
also possesses the SL(2,R) residual symmetry in the ``quasi-light-cone'' gauge.
In this sense, the correspondence between the SL(2,R) residual symmetry and
the area-preserving diffeomorphism invariance in both regularization
approaches does not exist. To quantise this theory, the string-like approach
is exploited, but a non-zero central charge in the Virasoro algebra appears,
and so the resulting quantum theory of gravity is anomalous, that is, at the
quantum level the area-preserving diffeomorphism invariance is broken.
However, this anomaly is only fictitious in the sense that it can be
avoided. In fact, in order to see that a consistent quantisation
of this theory is possible, a set of current-like fields is
introduced instead of the string-like variables, and the consistency condition
is the vanishing of the total central charge, which establishes the relation
between the constant {\em K} and the matter central charge $C_{m}$. Therefore, a
consistent quantum theory is obtained, which means that the area-preserving
diffeomorphism invariance can be maintained at the quantum level.

We start from the effective action with Weyl and area-preserving
diffeomorphism invariance [1-7]:
\begin{equation}
S=\int d^{2}x\left[-{\frac 1 2}{\gamma}^{{\mu}{\nu}}{\partial}_{\mu}{\phi}
{\partial}_{\nu}{\phi}+{\alpha}R(r){\phi}\right]
\end{equation}
where
\begin{equation}
{\gamma}^{{\mu}{\nu}} \equiv \sqrt{-g}g^{{\mu}{\nu}},
\end{equation}
is Weyl invariant and $\phi$ is an auxiliary field, and for simplicity in the following
discussion $\alpha=1$ is chosen. In fact, one may set the R$\phi$ coefficient
to be an arbitrary number {\em n} [6]. Following the ADM formulation,
the metric can be parametrized as [13-16]:
\begin{equation}
g_{{\alpha}{\beta}}=e^{2\rho}\left(
\begin{array}{cc}
-{\sigma}^2+{\theta}^2 & \theta \\
\theta & 1
\end{array}\right)
\end{equation}
where $\sigma(x)$ and $\theta(x)$ are lapse and shift functions respectively,
and the conformal factor $e^{{2}{\rho}}$ has been factored out. In terms of
this parametrization, the action (1) can be written as
\begin{eqnarray}
S=\int d^{2}x\left[\frac{{\dot{\phi}}^2}{2\sigma}-\frac{\theta}{\sigma}
\dot{\phi}{\phi}^{\prime}-\frac{{\sigma}^2-{\theta}^2}{2\sigma}{{\phi}
^{\prime}}^2+\frac{\dot{\phi}}{\sigma}\left(2{\theta}^{\prime}
-\frac{\theta{\sigma}^{\prime}}{\sigma}+\frac{\dot{\sigma}}{\sigma}\right)
\right. \nonumber \\
\left.+\frac{{\phi}^{\prime}}{\sigma}\left(\sigma{\sigma}^{\prime}
-2{\theta}{\theta}^{\prime}-\frac{\theta{\dot{\sigma}}}{\sigma}
+\frac{{\theta}^2{\sigma}^{\prime}}{\sigma}\right)\right]
\end{eqnarray}
which is independent of the conformal factor $\rho$, and so is Weyl-invariant.
In the above, dots and primes denote differentiation with respect to time
and space respectively. The canonical momenta associated with the fields
$\{\theta,\sigma,\phi \}$ are:
\begin{eqnarray}
{\pi}_{\theta} &\approx& 0 \\
{\pi}_{\sigma}&=&\frac{\dot{\phi}}{{\sigma}^2}-\frac{\theta{\phi}^{\prime}}
{{\sigma}^2}\nonumber \\
{\pi}_{\phi}&=&\frac{\dot{\phi}}{\sigma}
-\frac{\theta{\phi}^{\prime}}{\sigma}+\frac{1}{\sigma}\left(2{\theta}^{\prime}
-\frac{\theta{\sigma}^{\prime}}{\sigma}+\frac{\dot{\sigma}}{\sigma}\right)
\end{eqnarray}
with Poisson brackets
\begin{equation}
\{{\theta}(x),{\pi}_{\theta}(y)\}=
\{{\sigma}(x),{\pi}_{\sigma}(y)\}=
\{{\phi}(x),{\pi}_{\phi}(y)\}=\delta(x-y)
\end{equation}

As is usually done, Eq.(5) has to be taken as a primary constraint and it
holds only in a ``weak'' sense. The canonical Hamiltonian, up to surface terms,
is
\begin{equation}
H_c=\int dx\left[\theta(3{\sigma}^{\prime}{\pi}_{\sigma}+2\sigma{\pi}
^{\prime}_{\sigma}+{\pi}_{\phi}{\phi}^{\prime})
+{\sigma}^2{\pi}_{\sigma}{\pi}_{\phi}-{\frac 1 2}{\sigma}^3{\pi}^2
_{\sigma}+{\frac 1 2}\sigma({\phi}^{\prime})^2-{\sigma}^{\prime}{\phi}
^{\prime}\right]
\end{equation}
The consistency of the constraint ${\pi}_{\theta}{\approx}0$ under time
evolution requires that
\begin{equation}
G_{\theta}(x)=-\{{\pi}_{\theta}(x),H_c\}=
3{\sigma}^{\prime}{\pi}_{\sigma}+2\sigma{\pi}
^{\prime}_{\sigma}+{\pi}_{\phi}{\phi}^{\prime} \approx 0
\end{equation}
which is a secondary constraint, and satisfies
$$\{G_{\theta}(x),G_{\theta}(y)\}=[G_{\theta}(x)+G_{\theta}(y)]
{\partial}_x{\delta}(x-y)$$
According to the Dirac Hamiltonian procedure for constrained systems, the
Poisson brackets of the Hamiltonian with the constraints should vanish weakly ,
so we have
\begin{equation}
\{H_c,G_{\theta}(x)\}\approx [{\sigma}G_{\sigma}(x)]^{\prime}\approx 0
\end{equation}
with
\begin{equation}
G_{\sigma}(x)=\sigma{\pi}_{\sigma}{\pi}_{\phi}-{\frac 1 2}{\sigma}^2
{\pi}_{\sigma}^2+{\frac 1 2}{{\phi}^{\prime}}^2-\frac{{\sigma}^{\prime}
{\phi}^{\prime}}{\sigma}+\frac{2({\phi}^{\prime}\sigma)^{\prime}}{\sigma}
-\frac{c}{\sigma}\approx 0
\end{equation}
where {\em c} is an arbitrary constant, and the canonical Hamiltonian, up to
surface terms, turns out to be:
$$H_c=\int dx({\theta}G_{\theta}+{\sigma}G_{\sigma}+c)$$
>From Eq.(7), after a series of steps, we have
\begin{eqnarray}
\{G_{\sigma}(x),G_{\sigma}(y)\}&=&[G_{\theta}(x)+G_{\theta}(y)]
{\partial}_x{\delta}(x-y)\nonumber \\
\{G_{\theta}(x),G_{\sigma}(y)\}&=&[G_{\sigma}(x)+G_{\sigma}(y)]
{\partial}_x{\delta}(x-y)\nonumber \\
\{G_{\theta}(x),G_{\theta}(y)\}&=&[G_{\theta}(x)+G_{\theta}(y)]
{\partial}_x{\delta}(x-y)
\end{eqnarray}
Eq.(12) shows that $G_{\sigma}, G_{\theta}$ form the diffeomorphism algebra
under the Poisson bracket operation, that is, they are first-class constraints at the
classical level. However, one can see that ${G_{\sigma}, G_{\theta}}$ generate
the area-preserving diffeomorphism transformation, not the general diffeomorphism
transformation. It can be seen from Eq.(9), i.e.
$$\delta\phi (x)= -\{\int\epsilon (y)G_{\theta}(y)dy, \phi (x)\}\\
=\epsilon (x)\partial_x\phi (x),\\$$
that under the infinitesimal
transformation, the field $\phi$ behaves as a scalar field, but from Eq.(1), we
find that $\phi$ is not a scalar field under the reparametrization
transformation [2]. As claimed in Ref.[7], the field $\phi$ is a scalar field
only with respect to the area-preserving diffeomorphism transformation. So we
conclude that ${G_{\sigma}, G_{\theta}}$ are the generators of the
area-preserving diffeomorphism symmetry.

When the area-preserving diffeomorphism invariance is kept, i.e., no gauge
fixings are chosen for ${G_{\sigma}, G_{\theta}}$, we can just choose a gauge
fixing for the constraint ${\pi_{\theta}{\approx}}0$, then we find the usual
conformal gauge does not exist. However, there is the so-called
``quasi-light-cone'' gauge in which the reduced Lagrangian possesses the SL(2,R)
residual symmetry. To discuss the ``quasi-light-cone'' gauge, we need a field
redefinition which is given by
\begin{equation}
\sigma = (B+1)^{-1}
\end{equation}
then the ``quasi-light-cone'' can be defined by choosing the gauge fixing for
the constraint ${\pi}_{\theta}{\approx}0$ as
\begin{equation}
\theta =B(B+1)^{-1}
\end{equation}
i.e., the invariant line element is
$$
ds^2=-\frac{2e^{2\rho}}{B+1}\left[dx^{+}dx^{-}-B(dx^{+})^2\right]
$$
>From Eqs.(13,14), the action (4) is reduced to
\begin{equation}
S=\int d^{2}x \left[{\partial}_{+}{\phi}{\partial}_{-}{\phi}+B(
{\partial}_{-}{\phi})^2-2{\partial}_{-}B{\partial}_{-}{\phi}\right]
\end{equation}
where $x^{\pm}=(\frac{1}{\sqrt{2}})(x^{0}{\pm}x^1),
{\partial}_{\pm}=(\frac{1}{\sqrt{2}})({\partial}_{0}{\pm}{\partial}_1).$
It can be verified that the action (15) is invariant under the residual
symmetry transformations
\begin{eqnarray}
\delta{\phi}&=&\epsilon{\partial}_{-}{\phi}-{\partial}_{-}{\epsilon}\nonumber
\\
\delta{B}&=&-{\partial}_{+}{\epsilon}+{\epsilon}{\partial}_{-}B-B{\partial}_{-}
{\epsilon}\nonumber \\
{\partial}^{3}_{-}{\epsilon}&=&0
\end{eqnarray}
>From action (15), the momenta conjugate to $\phi$ and {\em B} are
\begin{eqnarray}
{\pi}_{\phi}&=&(1+B)(\dot{\phi}-{\phi}^{\prime})-{\phi}^{\prime}-\dot{B}
+B^{\prime}\nonumber \\
{\pi}_{B}&=&-\dot{\phi}+{\phi}^{\prime}
\end{eqnarray}
The transformation (16) is generated by the charge $Q=\int dx q(x)$ with
\begin{equation}
q(x)={\epsilon}\left[-\frac{B+1}{\sqrt{2}}T_{--}+\sqrt{2}{\partial}^{2}_{-}B
\right]
-\sqrt{2}({\partial}_{-}{\epsilon})({\partial}_{-}{B})+\sqrt{2}
({\partial}_{-}^{2}{\epsilon})B
\end{equation}
where $T_{--}$ is the $(--)$ component of the energy-momentum tensor given by
$$T_{--}=({\partial}_{-}{\phi})^2+2{\partial}_{-}^{2}{\phi}$$
If we make a decomposition of $\epsilon(x)$ with
\begin{equation}
{\epsilon}=\frac{1}{\sqrt{2}}[{\epsilon}^{-}(x^{+})+2x^{-}{\epsilon}^{0}(x^{+})
+(x^{-})^2{\epsilon}^{+}(x^{+})]
\end{equation}
the generator {\em q(x)} becomes
\begin{equation}
q(x)={\frac 1 2}{\epsilon}^{-}(x^{+})j^{+}-{\epsilon}^{0}(x^{+})j^{0}
+{\frac 1 2}{\epsilon}^{+}(x^{+})j^{-}
\end{equation}
with
\begin{eqnarray}
j^{+}&=&-(1+B)T_{--}+2{\partial}^{2}_{-}B \nonumber \\
j^{0}&=&\sqrt{2}{\partial}_{-}B-x^{-}j^{+}\nonumber \\
j^{-}&=&4B-2x^{-}j^{0}-(x^{-})^2j^{+}
\end{eqnarray}
Using the canonical commutation relations of ${\phi}$ and B, the currents (21)
obey the SL(2,R) algebra
\begin{equation}
\{j^{a}(x),j^{b}(y)\}=-2\sqrt{2}{\epsilon}^{abc}{\eta}_{cd}j^{d}(x){\delta}
(x-y)+4{\eta}^{ab}{\delta}^{\prime}(x-y)
\end{equation}
with
$${\eta}^{ab}=\left(
\begin{array}{ccc}
0 & 0 & 2 \\
0 & -1 & 0 \\
2 & 0 & 0
\end{array}\right), \qquad
{\epsilon}^{-0+}=1$$
where $a, b, c = -, 0, +$.  Eqs.(12,15,16,20,21) show that in the ``quasi-light-cone'' gauge, besides the
area-preserving diffeomorphism invariance, there is a SL(2,R) residual
symmetry. Then we conclude that the SL(2,R) residual symmetry in Polyakov's
light-cone approach [9] does not correspond to the area-preserving
diffeomorphism
symmetry in the Weyl-invariant approach [1-7].

    From the above discussion, we see that the SL(2,R) currents manifest
themselves as generators of the residual symmetry in the ``quasi-light-cone''
gauge, but these currents can be defined in a gauge-independent way by the
following expressions:
\begin{eqnarray}
j^{+}&=&{\sigma}(G_{\sigma}-G_{\theta})\nonumber \\
&=&{\sigma}^2{\pi}_{\sigma}{\pi}_{\phi}
-{\frac 1 2}{\sigma}^3{\pi}_{\sigma}^2+{\frac 1 2}{\sigma}{{\phi}^{\prime}}^2
+{\sigma}^{\prime}{\phi}^{\prime}
+2{\sigma}{\phi}^{{\prime}{\prime}}-3{\sigma}{\sigma}^{\prime}{\pi}_{\sigma}
-2{\sigma}^2{\pi}_{\sigma}^{\prime}-{\sigma}{\pi}_{\phi}{\phi}^{\prime}
\nonumber \\
j^{0}&=&\sqrt{2}\left(-{\sigma}{\pi}_{\sigma}+{\pi}_{\phi}+\frac{{\sigma}
^{\prime}}{\sigma}\right)-x^{-}j^{+}\nonumber \\
j^{-}&=&4\left(\frac{1}{\sigma}-1\right)-2x^{-}j^{0}-(x^{-})^2j^{+}
\end{eqnarray}
For simplicity, the arbitrary constant {\em c} in $G_{\sigma}$ has been chosen
to be zero from now on. One can easily check that the currents defined above
satisfy Eq.(22), and Eq.(23) can be reduced to Eq.(21) in the
``quasi-light-cone'' gauge.

   From Eq.(23) one can obtain
\begin{equation}
{\sigma}=\left\{1+{\frac 1 4}\left[j^{-}+2x^{-}j^{0}+(x^{-})^2j^{+}\right]
\right\}^{-1}
\end{equation}
which shows that the graviton field can be expressed in terms of the {\em j}
variables.

In order to quantise this theory by the string-like approach, we perform the
canonical change of the original variables by
$${\psi}={\phi}+\ln{\sigma}, \qquad
{\eta}=-\ln{\sigma}$$
and
\begin{equation}
{\pi}_{\psi}={\pi}_{\phi}, \qquad
{\pi}_{\eta}={\pi}_{\phi}-{\pi}_{\sigma}e^{-\eta}
\end{equation}
In terms of the new canonical variables defined in (25), the area-preserving
diffeomorphism constraints can be written as
\begin{eqnarray}
G_{\sigma}&=&{\frac 1 2}{{\psi}^{\prime}}^2+{\frac 1 2}{\pi}_{\psi}^{2}
+2{\psi}^{{\prime}{\prime}}-{\frac 1 2}{{\eta}^{\prime}}^2-{\frac 1 2}
{\pi}_{\eta}^{2}+2{\eta}^{{\prime}{\prime}}\nonumber \\
G_{\theta}&=&{\pi}_{\psi}{\psi}^{\prime}+2{\pi}_{\psi}^{\prime}+{\pi}_{\eta}
{\eta}^{\prime}-2{\pi}_{\eta}^{\prime}
\end{eqnarray}
which are equivalent to the constraints
\begin{equation}
G_{\pm}={\frac 1 2}(G_{\sigma}{\pm}G_{\theta})
\end{equation}
which obey the Virasoro algebra
\begin{eqnarray*}
\{G_{\pm}(x),G_{\pm}(y)\}&=&\mp[G_{\pm}(x)+G_{\pm}(y)]{\partial}_{x}
{\delta}(x-y)\nonumber \\
\{G_{+}(x),G_{-}(y)\}&=&0
\end{eqnarray*}
When we take the Fourier transform of the constraint $G_{+}$
$$L_n=\int_{-\pi}^{\pi}dx\exp(inx)G_{+}(x)$$
we have
\begin{equation}
\{L_n,L_m\}=i(n-m)L_{n+m}
\end{equation}
If we carry on the analogy with string theory by defining the oscillator
variables
\begin{eqnarray}
a_n&=&\int_{-\pi}^{\pi}dx\exp(inx)({\pi}_{\psi}+{\psi}^{\prime})\nonumber \\
b_n&=&\int_{-\pi}^{\pi}dx\exp(inx)({\pi}_{\eta}-{\eta}^{\prime})
\end{eqnarray}
which satisfy
\begin{equation}
\{a_n,a_m\}=-\{b_n,b_m\}=-2in{\delta}_{n,-m}
\end{equation}
then the Virasoro operators have the familiar form
$$L_n={\frac 1 2}\sum\limits_{m}a_{n-m}a_m-{\frac 1 2}\sum\limits_{m}b_{n-m}b_m
-in(a_n-b_n)$$
Considering the induced 2D-gravity with Weyl and area-preserving
diffeomorphism invariance described by the two sets of operators
$a_{n}$ and $b_{n}$ with commutation relations taken from (30), we can
define the quantum Virasoro generators using normal ordering
\begin{equation}
L_n={\frac 1 2}\sum\limits_{m}:a_{n-m}a_m:
-{\frac 1 2}\sum\limits_{m}:b_{n-m}b_m:-in(a_n-b_n)
\end{equation}
and the vacuum state $|0>$ is defined by
\begin{equation}
L_{n}|0>=0, \qquad n>0, \qquad L_{0}|0>=a_{0}|0>
\end{equation}
If such a prescription is followed, a non-zero central
charge in the Virasoro algebra appears [17-19], and the resulting quantum theory
of gravity is anomalous.

However, if the string-like variables $\psi$ and $\eta$ are replaced by a set of current-like
fields, a consistent quantum theory can be achieved. First we
introduce a new field {\em h} given by
\begin{equation}
h={\pi}_{\phi}+{\phi}^{\prime}+\frac{{\sigma}^{\prime}}{\sigma}
\end{equation}
with
\begin{eqnarray*}
\{h(x),h(y)\}&=&2{\partial}_{x}{\delta}(x-y)\\
\{h(x),j^{a}(y)\}&=&0 ,       a = -, 0, +
\end{eqnarray*}
Eq.(24) shows that the field $\sigma$ can be expressed in terms of the {\em j}
variables, while ${\pi}_{\sigma}, {\pi}_{\phi}, {\phi}$ can be solved using (23)
and (33) with appropriate boundary conditions. Since the set of variables
$(j^{-}, j^{0}, j^{+}, h)$ is defined in a gauge-independent way, we then
describe the whole theory in terms of them, and the constraints can be
expressed as
\begin{eqnarray*}
G_{\theta}&=&G_m+G_s-\frac{x^{-}}{\sqrt{2}}{j^{+}}^{\prime}\approx 0 \\
G_{\sigma}&=&G_{\theta}+\frac{j^{+}}{\sigma}\approx 0
\end{eqnarray*}
with
\begin{eqnarray}
G_m&=&{\frac 1 4}h^2+h^{\prime}\nonumber \\
G_s&=&{\frac 1 8}{\eta}_{ab}j^{a}j^{b}-\frac{1}{\sqrt{2}}{j^{0}}^{\prime}
\end{eqnarray}
>From (34), we find the old constraints $(G_{\theta},G_{\sigma})$ are
equivalent to
\begin{eqnarray}
&j^{+}\approx 0 \nonumber \\
&b=G_m+G_s \approx 0
\end{eqnarray}
which obey the algebra
\begin{eqnarray}
\{j^{+}(x),j^{+}(y)\}&=&0\nonumber \\
\{b(x),j^{+}(y)\}&=&-[{\partial}_{x}j^{+}(x)]{\delta}(x-y)\nonumber \\
\{b(x),b(y)\}&=&[b(x)+b(y)]{\partial}_{x}{\delta}(x-y)
\end{eqnarray}
Consider the algebra of $G_{m}$ (where the {\em h} field enters), we have
\begin{eqnarray}
\{G_m(x),G_m(y)\}&=&[G_m(x)+G_m(y)]{\delta}^{\prime}(x-y)-2{\delta}^{
{\prime}{\prime}{\prime}}(x-y)\nonumber \\
\{G_m(x),G_s(y)\}&=&0
\end{eqnarray}
which shows that $G_{m}$ retains some memory of the matter fields through the
semi-classical central charge in (37). Therefore we can consider the
{\em h} field as
carrying the matter central charge in this representation of the theory [17].
According to the light-cone gauge prescription [9], we assume the $j^{a}$
operators satisfy equal-time commutation relations as in (22),
\begin{equation}
[j^{a}(x),j^{b}(y)]=-2i\sqrt{2}{\epsilon}^{abc}{\eta}_{cd}j^{d}{\delta}(x-y)
+4i{\bar{\alpha}}^{2}{\eta}^{ab}{\delta}^{\prime}(x-y)
\end{equation}
allowing a possible renormalization of the constant $\alpha$ in action (1), and
the renormalized constant $\bar{\alpha}$ is {\em K/384}, where {\em K} is to be
determined by the consistency condition. Then the quantum constraint operators
can be defined as
\begin{eqnarray}
&j^{+}\approx 0 \nonumber \\
&b=:G_m:+:G_g:+:G_s:\approx 0
\end{eqnarray}
where
$$:G_s:=\frac{48\pi}{K+2}:{\eta}_{ab}j^{a}j^{b}:-\frac{1}{\sqrt{2}}
{j^{0}}^{\prime}\approx 0$$
is the renormalized gravitational energy-momentum contribution and $G_{g}$ is
the ghost piece coming from the gauge fixing. The usual physical state can be
defined as
\begin{equation}
j^{+}|physical>=0, \qquad
b|physical>=0
\end{equation}
Eq.(40) demands that the algebra of the constraints $(j^{+},b)$ has no
Schwinger terms. Then the consistency condition is the vanishing of the total
central charge of $[b,b]$ [17]
\begin{equation}
C_T=C_m+C_g+C_s=0
\end{equation}
with
\begin{equation}
C_s=3K/(K+2)-6K
\end{equation}
which establishes the relation between {\em K} and the matter central charge
$C_{m}$. Then a consistent quantum theory can be obtained, that is, at the
quantum level the area-preserving diffeomorphism invariance has been
maintained.

In conclusion, the constraint structure for the induced 2D-gravity with the
Weyl and area-preserving diffeomorphism invariance has been analysed
completely in the ADM formulation. It has been shown that if the
area-preserving diffeomorphism constraints are kept, the usual conformal gauge
does not exist, whereas we can choose the ``quasi-light-cone'' gauge. It has
been first found that in the ``quasi-light-cone'' gauge, besides the
area-preserving diffeomorphism invariance, there is also a SL(2,R) residual
symmetry in the reduced Lagrangian. In this sense, the correspondence between
the SL(2,R) residual symmetry and the area-preserving diffeomorphism
invariance in both regularisation approaches does not hold. Although the
SL(2,R) currents manifest themselves as generators of the residual symmetry in
the ``quasi-light-cone'' gauge, we have found that these currents can be defined
in a gauge-independent way. When the string-like approach is applied to
quantise this theory, a non-zero central charge in the Virasoro algebra
appears, and the resulting quantum theory of gravity is anomalous. In
order to consistently quantize this theory, a set of the SL(2,R) current-like
fields have been introduced instead of the string-like variables, and the
consistency condition is the vanishing of the total central charge. Then a
consistent quantum theory is obtained, which means that the
area-preserving diffeomorphism invariance can be maintained at the quantum
level.
\vskip 20mm
\noindent
{\bf Acknowledgement}
\par
This work was supported in part by the European Union under the Human Capital
and Mobility Programme. J.-G. Z. acknowledges support of the Alexander von Humboldt Foundation
in the form of a research fellowship and J.-Q.L. support of the Deutsche
Forschungsgemeinschaft.

\end{document}